\documentstyle[twoside,fleqn,espcrc2]{article}

\input epsf

\def\rmd{{\rm d}}

\def\rme{{\rm e}}

\def\proof{\noindent{\sl Proof:}\kern0.6em}

\def\frac#1#2{\hbox{$#1\over#2$}}
\def\dual{\mathstrut^*\kern-0.1em}

\def\lvec#1{\setbox0=\hbox{$#1$}
    \setbox1=\hbox{$\scriptstyle\leftarrow$}
    #1\kern-\wd0\smash{
    \raise\ht0\hbox{$\raise1pt\hbox{$\scriptstyle\leftarrow$}$}}
    \kern-\wd1\kern\wd0}
\def\rvec#1{\setbox0=\hbox{$#1$}
    \setbox1=\hbox{$\scriptstyle\rightarrow$}
    #1\kern-\wd0\smash{
    \raise\ht0\hbox{$\raise1pt\hbox{$\scriptstyle\rightarrow$}$}}
    \kern-\wd1\kern\wd0}

\def\nabstar#1{\nabla\kern-0.5pt\smash{\raise 4.5pt\hbox{$\ast$}}
               \kern-4.5pt_{#1}}

\def\drvstar#1{\partial\kern-0.5pt\smash{\raise 4.5pt\hbox{$\ast$}}
               \kern-5.0pt_{#1}}

\def\MeV{{\rm MeV}}

\def\rhoprime{\rho\kern1pt'}
\def\rhobar{\bar{\rho}}
\def\rhobarprime{\rhobar\kern1pt'}
\def\rhobartilde{\kern2pt\tilde{\kern-2pt\rhobar}}
\def\rhobartildeprime{\kern2pt\tilde{\kern-2pt\rhobar}\kern1pt'}

\def\zetabar{\bar{\zeta}}
\def\zetaprime{\zeta\kern1pt'}
\def\zetabarprime{\zetabar\kern1pt'}
\def\zetar{\zeta_{\raise-1pt\hbox{\sixrm R}}}
\def\zetabarr{\zetabar_{\raise-1pt\hbox{\sixrm R}}}

\def\phiimpr{\phi_{\kern0.5pt\hbox{\sixrm I}}}

\def\dirac#1{\gamma_{#1}}
\def\diracstar#1#2{
    \setbox0=\hbox{$\gamma$}\setbox1=\hbox{$\gamma_{#1}$}
    \gamma_{#1}\kern-\wd1\kern\wd0
    \smash{\raise4.5pt\hbox{$\scriptstyle#2$}}}

\def\ba{b_{\rm A}}
\def\bp{b_{\rm P}}

\def\fp{f_{\rm P}}

\def\opprime#1{\setbox0=\hbox{${\cal O}$}\setbox1=\hbox{${\cal O}_{\rm #1}$}
    {\cal O}_{\rm #1}\kern-\wd1\kern\wd0
    \smash{\raise4.5pt\hbox{\kern1pt$\scriptstyle\prime$}}\kern1pt}

\def\ophatprime#1{\setbox0=\hbox{$\widehat{\cal O}$}
    \setbox1=\hbox{$\widehat{\cal O}_{\rm #1}$}
    \widehat{\cal O}_{\rm #1}\kern-\wd1\kern\wd0
    \smash{\raise4.5pt\hbox{\kern1pt$\scriptstyle\prime$}}\kern1pt}

\def\bopprime#1{\setbox0=\hbox{${\cal O}$}\setbox1=\hbox{${\cal O}_{\rm #1}$}
    {\cal L}_{\rm #1}\kern-\wd1\kern\wd0
    \smash{\raise4.5pt\hbox{\kern1pt$\scriptstyle\prime$}}\kern1pt}

\def\blagprime#1{\setbox0=\hbox{${\cal B}$}\setbox1=\hbox{${\cal B}_{#1}$}
    {\cal B}_{#1}\kern-\wd1\kern\wd0
    \smash{\raise5.2pt\hbox{\kern1pt$\scriptstyle\prime$}}\kern1pt}

\def\gbar{\bar{g}}

\def\mq{m_{\rm q}}

\def\mbar{\kern1pt\overline{\kern-1pt m\kern-1pt}\kern1pt}

\def\za{Z_{\rm A}}

\def\zp{Z_{\rm P}}
\def\zpmom{\zp^{\raise1pt\hbox{\sixrm MOM}}}

\def\msbar{{\rm \overline{MS\kern-0.05em}\kern0.05em}}
\def\lat{{\rm lat}}

\def\Lmax{L_{\rm max}}

\title{\vspace{-4.05cm}
       \rightline{\normalsize DESY 97-188}
       \vspace{3.5cm}
       Non-perturbative quark mass renormalization%
       \thanks{Talk given by M.L.~at
       the International Symposium on
       Lattice Field Theory, July~22--26, 1997, 
       Edinburgh}}

\author{S.~Capitani\address{Deutsches 
        Elektronen-Synchrotron DESY,
        Notkestrasse 85, D-22603 Hamburg, Germany},
        M.~Guagnelli\address{CERN, Theory Division,
        CH-1211 Gen\`eve 23, Switzerland},
        \addtocounter{address}{-2}
        M.~L\"uscher\addressmark,
        \addtocounter{address}{+1}
        S.~Sint\address{SCRI, The Florida State University, 
        Tallahassee, FL 32306-4130, USA},
        R.~Sommer\address{DESY-IfH Zeuthen, Platanenallee 6, 
        D-15738 Zeuthen, Germany},
        P.~Weisz\address{Max-Planck-Institut f\"ur Physik,
        F\"ohringerring 6, D-80805 M\"unchen, Germany}\hspace{1ex}and
        H.~Wittig\address{Theoretical Physics, University of Oxford,
        1 Keble Road, Oxford OX1 3NP, England}}

\begin{document}

\begin{abstract}
We show that the renormalization factor relating the 
renormalization group invariant quark masses to the bare quark masses 
computed in lattice QCD can be determined non-perturbatively.
The calculation is based on an extension of a finite-size technique 
previously employed to compute the running coupling in quenched QCD.
As a by-product we obtain the $\Lambda$--parameter in this theory 
with completely controlled errors.
\end{abstract}

\maketitle

\section{INTRODUCTION}

Calculations of the quark masses in lattice QCD are in principle
straightforward, but there are several sources of systematic errors
which must be carefully studied (see ref.~\cite{BhattacharyaGupta} for a
review of the status of these calculations and an up-to-date list of
references). One of the uncertainties arises from the renormalization
constant needed to convert from the lattice normalizations to the
$\msbar$ scheme of dimensional regularization. Usually one relies on
bare perturbation theory (or some modified form thereof) to evaluate
this factor. Since only the one-loop term of the expansion is known,
and since the gauge coupling is not small on the accessible lattices,
the associated error is, however, difficult to estimate.
While the extension of the perturbation expansion to the next order may be
helpful at this point, it is quite clear that a non-perturbative
determination of the renormalization factor will be required 
to remove all doubts on the reliability of the quark mass calculations
in lattice QCD.

The calculation of the quark mass renormalization factor discussed in
this talk is based on a recursive finite-size technique which allows
one to compute the scale evolution of the renormalized parameters and
fields from low to very high energies. An uncontrolled application of
perturbation theory can thus be avoided. The general strategy of the
computation has already been described in ref.~\cite{letter} and it is
our aim here to report on the progress that has been made along these
lines in the case of the quark mass renormalization
(for an alternative approach to the problem 
see refs.~\cite{MartinelliEtAl,GiustiEtAl}).

\section{PCAC RELATION}

Quark mass ratios can be accurately estimated using 
chiral perturbation theory 
(see ref.~\cite{Leutwyler} for a recent discussion). 
To determine the absolute values
of the quark masses in any given renormalization scheme
it is hence sufficient to compute a particular linear combination
of them such as the sum of the up quark mass and the strange quark mass.
A possible starting point then is the PCAC relation
\begin{equation}
  \partial_{\mu}(\bar{u}\dirac{\mu}\dirac{5}s)_{\rm R}=
  (\mbar_{\rm u}+\mbar_{\rm s})(\bar{u}\dirac{5}s)_{\rm R}
\end{equation}
between the renormalized $\Delta S=1$ axial current and the associated
renormalized pseudo-scalar density. 

In O($a$) improved lattice QCD the axial 
current and density are given by \cite{paperI}
\begin{equation}
  (\bar{u}\dirac{\mu}\dirac{5}s)_{\rm R}=
  \za
  \left(1+\ba a\mq\right)
  (\bar{u}\dirac{\mu}\dirac{5}s)_{\lat},
\end{equation}

\begin{equation}
  (\bar{u}\dirac{5}s)_{\rm R}=
  \zp
  \left(1+\bp a\mq\right)
  (\bar{u}\dirac{5}s)_{\lat}, 
\end{equation} 
where $(\ldots)_{\lat}$
denotes the unrenormalized improved fields and $\mq$ is
the average of the bare $u$ and $s$ quark masses (with the additive
renormalization taken into account). 
Eq.~(1) then holds up to terms of order $a^2$. 
In the present context the factors $1+b_{\rm X}a\mq$ amount to
corrections of a few percent at most.
The one-loop expression \cite{WeiszSint}
\begin{equation}
    \ba-\bp=-0.001\times g_0^2+\ldots
\end{equation}
and a recent non-perturbative calculation 
of this difference \cite{PetronzioEtAl} moreover
suggest that these factors nearly cancel in eq.~(1).
At the present level of accuracy it thus appears safe to drop them.

So if we determine $(m_{\rm u}+m_{\rm s})_{\lat}$ through
the vacuum-to-kaon matrix element of the unrenormalized PCAC relation,
\begin{equation}
  \partial_{\mu}(\bar{u}\dirac{\mu}\dirac{5}s)_{\lat}=
  (m_{\rm u}+m_{\rm s})_{\lat}(\bar{u}\dirac{5}s)_{\lat},
\end{equation}
it follows that 
\begin{equation}
  \mbar_{\rm u}+\mbar_{\rm s}=
  (m_{\rm u}+m_{\rm s})_{\lat}\za/\zp.
\end{equation}
The calculation of $(m_{\rm u}+m_{\rm s})_{\lat}$ is
standard by now \cite{BhattacharyaGupta} and will not be discussed here.
As for the renormalization factors we note that 
$\za(g_0)$ has been computed non-perturbatively \cite{paperIV}
to a precision of about $1\%$, using
a variant of the chiral Ward identity method of 
refs.~\cite{BochicchioEtAl}--\cite{HentyEtAl}.
 
The remaining unknown factor in eq.~(6) thus is the 
renormalization constant $\zp(g_0,a\mu)$. 
This factor is much harder to determine than $\za(g_0)$, 
because it is a function of two variables, the gauge coupling 
and the normalization scale $\mu$ of the chosen renormalization scheme.

\begin{figure}
\vspace{-3.4cm}
\begin{center}
\hbox{\kern-0.8cm\epsfxsize=9.6cm\epsfbox{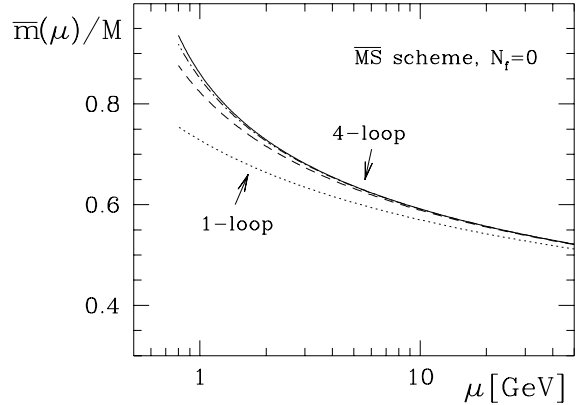}}
\end{center}
\vspace{-2.5cm}
\caption{Scale evolution of the running quark masses in 
         quenched QCD}
\vspace{-0.5cm}
\end{figure}

\section{SCALE DEPENDENCE}

In the continuum theory the scale evolution of the 
running coupling and quark masses is
governed by the renormalization group equations
\begin{equation}
  \mu{\partial\gbar\over\partial\mu}=
  \beta(\gbar),
  \qquad
  \mu{\partial\mbar\over\partial\mu}=
  \tau(\gbar)\mbar.
\end{equation}
For the case of dimensional regularization with minimal subtraction the 
$\beta$- and $\tau$-functions have been calculated up to 4-loop order
of perturbation theory 
(\cite{RitbergenI}--\cite{RitbergenII} and references cited there).
Using these results the evolution equations
can be integrated starting at some large $\mu$ where the 
value of the running coupling is known.
If we define the renormalization group invariant quark masses $M$ through
\begin{equation}
  M=\lim_{\mu\to\infty}
  \mbar\,(2b_0\gbar^2)^{-d_0/2b_0},
\end{equation}
where $b_0$ and $d_0$ are the one-loop coefficients of the $\beta$- and 
the $\tau$-function respectively, the curves shown in fig.~1 
are thus obtained. 

\begin{figure}
\vspace{0.1cm}
\begin{center}
\hbox{\kern0.3cm\epsfxsize=4.8cm\epsfbox{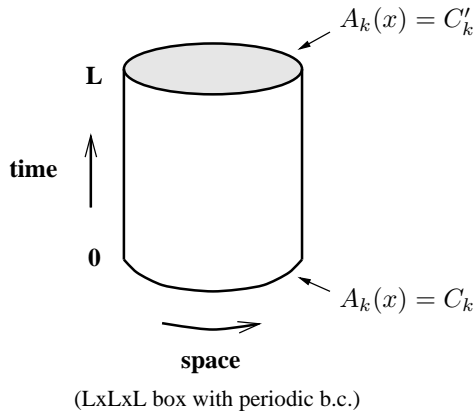}}
\end{center}
\vspace{-6.3cm}
\hbox{\kern4.7cm$A_k(x)=C'_k$}
\vspace{3.3cm}
\hbox{\kern4.7cm$A_k(x)=C_k$}
\vspace{0.8cm}
\caption{Space-time manifold used to set up the SF scheme}
\vspace{-0.7cm}
\end{figure}

It should be evident at this point that the knowledge of the 
running quark masses at high energies is equivalent
to providing the values of the corresponding renormalization group
invariant quark masses. In other words, instead of
$\mbar_{\rm u}+\mbar_{\rm s}$ we may just as well compute 
$M_{\rm u}+M_{\rm s}$. It is easy to show that the 
renormalization group invariant masses are 
non-perturbatively defined and that they do not depend
on the chosen renormalization scheme. 
We can hence compute them in any scheme that we like
such as the one described in the following paragraphs.

\section{SF SCHEME}

The Schr\"odinger functional (SF) scheme is a particular
finite-volume renormalization scheme which has
previously been used to compute the running coupling in the pure
SU(3) gauge theory \cite{alphaI}. Its application in the present context
has already been discussed in ref.~\cite{letter} and in this
preliminary report we only describe some
of the key features of the scheme. 

The basic idea is to consider QCD in a finite space-time volume of
size $L$ in all directions. Renormalization conditions are then
specified at scale $\mu=1/L$ and vanishing quark masses. To be able to
perform numerical simulations at zero quark masses the
boundary conditions are chosen 
in such a way that a frequency gap of order $1/L$
is induced on the quark and gluon fields. This can be achieved by assuming
the space-time mani\-fold to be as shown in fig.~2 and by imposing
Dirichlet boundary conditions at time $x_0=0$ and $x_0=L$.
More precisely, the spatial components of the 
gauge field are required to be equal to some prescribed
constant abelian fields $C$ and $C'$. 
The response of the system to a change of these boundary values
may be used to define a running coupling $\gbar^2$. We do not
give any further details here, because the definition of $\gbar^2$
that we have used is exactly the same as in ref.~\cite{alphaI}.

\begin{figure}
\vspace{-0.2cm}
\begin{center}
\hbox{\kern1.2cm\epsfxsize=4.8cm\epsfbox{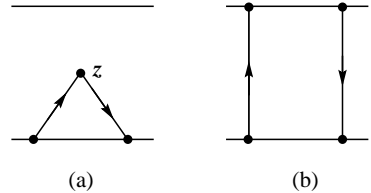}}
\end{center}
\vspace{-1.3cm}
\caption{Quark diagrams contributing to $\fp(z)$ (dia\-gram a) and
$f_1$ (diagram b). The horizontal lines indicate the boundaries
of space-time at $x_0=0$ and $x_0=L$. The pseudo-scalar density
is inserted at the point $z$}  
\vspace{-0.5cm}
\end{figure}

The renormalized quark masses in the SF scheme are defined
through the PCAC relation and the renormalization condition 
for the pseudo-scalar density (cf.~sect.~2).
The precise form of the latter is of only practical importance
since all definitions lead to the same values of the 
renormalization group invariant quark masses.
A simple choice for the renormalization constant is
\begin{equation}
  \zp(g_0,L/a)=c\sqrt{f_1}/\fp(z)|_{z_0=L/2},
\end{equation}
where $\fp(z)$ and $f_1$ are unrenormalized correlation functions 
involving the pseudo-scalar density and the quark fields
at the boundaries of space-time
\cite{paperI,paperIV} ($c$ is to be chosen such that
$\zp=1$ at $g_0=0$). 
The corresponding
quark diagrams are shown in fig.~3. Note that the renormalization
factors associated with the boundary quark fields cancel in the ratio (9). 
In particular, the renormalized coupling and quark masses 
defined in this way satisfy the usual renormalization group equations 
[eq.~(7)] with $\mu=1/L$ and the appropriate $\beta$- and $\tau$-functions. 

\begin{figure}
\vspace{-0.6cm}
\begin{center}
\hbox{\kern-0.2cm\epsfxsize=8.0cm\epsfbox{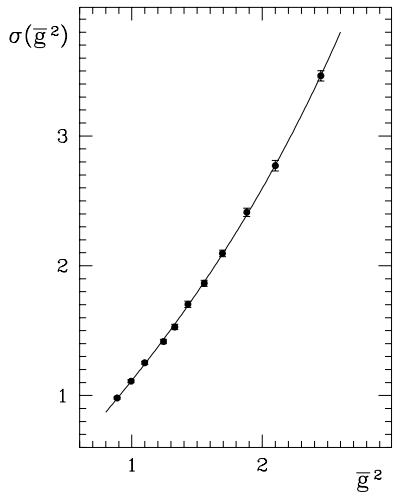}}
\end{center}
\vspace{-2.5cm}
\caption{Simulation results for the 
step scaling function $\sigma(\gbar^2)$
(quenched QCD)}
\vspace{-0.7cm}
\end{figure}

\section{RENORMALIZATION GROUP}

In the SF scheme
a change of the normalization scale amounts to 
a change of the lattice size at fixed bare parameters. 
By considering pairs of lattices with sizes $L$ and $2L$, we can
thus study the evolution of the 
running coupling and quark masses under changes of the normalization
scale by factors of $2$. The important point to note is that 
the box size $L$ can be as small as we like in physical units.
The only restriction is that $L$ and the low-energy physical scales
in the theory (such as the radius $r_0$ \cite{rnull}) 
should be significantly greater than the lattice spacing to avoid
large cutoff effects. Such studies can be carried out
using numerical simulations and the scale evolution of the 
renormalized parameters is thus obtained non-perturbatively.

Close to the continuum limit
the evolution from size $L$ to $2L$ is described
by the ``step scaling functions" $\sigma$ and $\sigma_{\rm P}$
through
\begin{equation}
  \gbar^2(2L)=\sigma(\gbar^2(L)),
\end{equation}

\begin{equation}
  \zp(2L)=\sigma_{\rm P}(\gbar^2(L))\zp(L).
\end{equation}
It is our experience that 
the lattice corrections to these equations 
are small and can be extrapolated away by repeating the computation
of the step scaling functions for various lattice spacings at fixed
$\gbar^2$. 
The step scaling function associated with the coupling 
has first been calculated in ref.~\cite{alphaI} and further data 
have since then been added so that this function 
is now known very accurately over a large range of couplings
(see fig.~4). 
As for the other function, $\sigma_{\rm P}$,
the available data (fig.~5) are already quite usable and 
further runs will be made to fill the gap 
at large couplings and to achieve a better precision.
All these results
refer to the continuum limit, i.e.~they involve an extrapolation
to $a=0$ as indicated above.

\begin{figure}
\vspace{-2.80cm}
\begin{center}
\hbox{\kern-0.6cm\epsfxsize=8.0cm\epsfbox{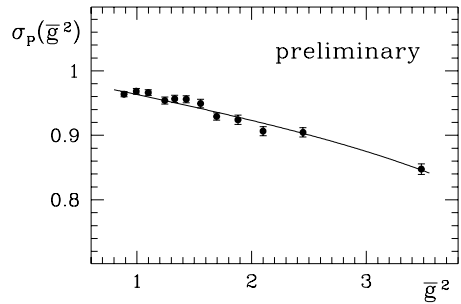}}
\end{center}
\vspace{-2.7cm}
\caption{Preliminary results for the step scaling function 
$\sigma_{\rm P}(\gbar^2)$ (quenched QCD)}
\vspace{-0.5cm}
\end{figure}

Once the step scaling functions have been determined, the sequence 
of couplings and quark masses
\begin{equation}
  u_k=\gbar^2(2^kL),\qquad
  m_k=\mbar(2^kL),
\end{equation}
is obtained recursively using
\begin{equation}
  u_{k+1}=\sigma(u_k),\qquad
  m_{k+1}=m_k/\sigma_{\rm P}(u_k).
\end{equation}
Since $\sigma(u)$ is monotonic 
the recursion can also be solved backwards, i.e.~we can move up and down
the energy scale as we wish.

A technical problem here is that 
the step scaling functions are only known 
at certain values of the coupling and only to a finite numerical precision.
This difficulty can be resolved by fitting the data with a
polynomial, as shown in the figures,
and using the fit functions in the recursion (13).
We have verified that the systematic error which is incurred by this
procedure is neglible compared to the statistical errors 
provided one stays in the range of couplings covered by the data.

\section{RESULTS}

The largest value of $\gbar^2$ which can be reached with the available data
for the step scaling functions is $3.48$. This corresponds to a certain
box size $\Lmax$, which is large enough that contact can be made
with the physical low-energy scales in the theory. 
Following ref.~\cite{alphaI}, and using some recent results
of the UKQCD collaboration \cite{WittigI,WittigII},
one finds $\Lmax/r_0=0.680(26)$. This can be taken as  
initial condition for the non-perturbative evolution of the coupling
and the recursion (13) then yields
the data points shown in fig.~6. 
For comparison we also plot the curves that one obtains
by integrating the 2- and 3-loop evolution
equation starting at the right-most data point.
Evidently the exact evolution of the SF coupling
is accurately matched by perturbation theory
down to surprisingly low energies.

\begin{figure}
\vspace{-2.0cm}
\begin{center}
\hbox{\kern-0.7cm\epsfxsize=9.0cm\epsfbox{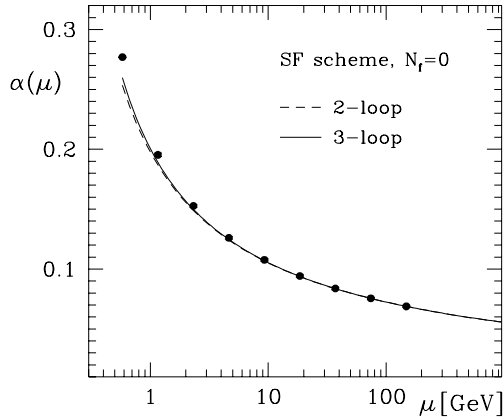}}
\end{center}
\vspace{-2.9cm}
\caption{Evolution of the running coupling $\alpha(\mu)=\gbar^2/4\pi$,
$\mu=1/L$, in quenched QCD, taking $r_0=0.5$ fm to convert to physical units}
\vspace{-0.5cm}
\end{figure}

We now digress a little bit and discuss how the 
$\Lambda$-parameter can be extracted from the 
calculated values of the running coupling.
The idea is to evaluate the exact expression
\begin{displaymath}
  \Lambda=\mu(b_0\gbar^2)^{-b_1/2b_0^2}\rme^{-1/2b_0\gbar^2}
\end{displaymath}
\begin{equation}
  \kern1.5em\times\exp\left\{-\int_0^{\gbar}\rmd g
  \left[{1\over\beta(g)}
  +{1\over b_0g^3}-{b_1\over b_0^2g}\right]\right\}
\end{equation}
at the high-energy data points of fig.~6. 
In this range of couplings the integral in the last factor
may be reliably calculated using the
perturbation expansion of the $\beta$-function,
which (in the SF scheme) is now known to 3-loop order \cite{threeloop}.
Higher-order corrections are negligible at this point,
and after converting to the $\msbar$ scheme the result
\begin{equation}
  \Lambda^{(0)}_{\msbar}=251\pm21\,\MeV
\end{equation}
is obtained,
where $r_0=0.5$ fm has been used to set the scale 
(the index $\raise1pt\hbox{$\scriptstyle(0)$}$ 
reminds us that this number is for quenched QCD).

\begin{figure}
\vspace{-2.0cm}
\begin{center}
\hbox{\kern-0.4cm\epsfxsize=9.0cm\epsfbox{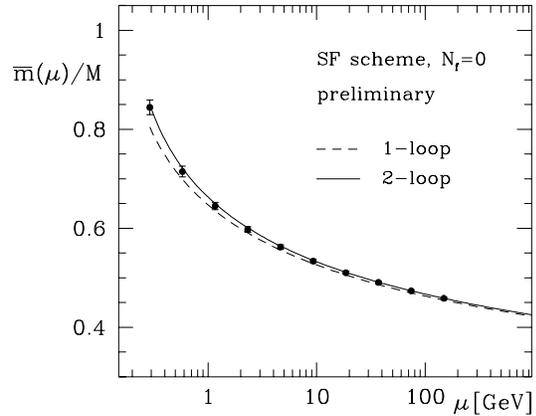}}
\end{center}
\vspace{-2.9cm}
\caption{Evolution of the running quark mass given in 
units of the renormalization group invariant mass $M$ [eq.~(8)]}
\vspace{-0.5cm}
\end{figure}

Now that the evolution of the coupling is under control, it is
straightforward to obtain the data points in fig.~7 by solving
the quark mass part of the recursion (13). Initially the recursion 
yields the ratio $\mbar(L)/\mbar(2\Lmax)$. As in the case of the 
$\Lambda$ parameter, the evolution may then be continued to 
infinite energy using the 2-loop expression for the $\tau$-function in
the SF scheme. In this way one is able to extract 
the renormalization group invariant mass $M$ and hence 
the ratio $\mbar(L)/M$ plotted in fig.~7. Note that here too
the scale evolution is accurately reproduced by perturbation theory
down to the lowest energies. In particular, there is 
little doubt that the use of the 
perturbative evolution at high energies is safe. 
We would like to emphasize, however, that the absence of 
large corrections to the perturbative evolution 
of the running quark mass depends on the chosen scheme
and should not be taken as a general feature of the theory.

At the lowest-energy data point in fig.~7 we have
\begin{equation}
  L=2\Lmax, \qquad M/\mbar=1.18(2).
\end{equation}
Recalling our discussion in sects.~2 and 3, it follows from this that 
the total renormalization factor, relating the bare current quark mass
appearing in eq.~(5) to the renormalization group invariant mass, is 
given by
\begin{equation}
  M/m_{\lat}=1.18(2)\times\za(g_0)/\zp(g_0,2\Lmax/a).
\end{equation}
The factors $\za$ and $\zp$ 
on the right-hand side of this equation
are listed in table~1 for two values of the bare coupling $\beta=6/g_0^2$.
Taking the product then yields the desired renormalization factor.
These numbers are still preliminary, but they show 
what can be expected to come out once our study has been completed.

\begin{table}
\vspace{0.0cm}
\setlength{\tabcolsep}{1.3ex}
\caption{Evaluation of eq.~(17)}
\begin{tabular}{ c  c  c  c  c }
\noalign{\vskip0.5ex}
\hline
\noalign{\vskip0.5ex}
$\beta$ & $2\Lmax/a$ & $\zp$ & $\za$ & $M/m_{\lat}$ \\
\noalign{\vskip0.5ex}
\hline
\noalign{\vskip0.5ex}
{$6.0$} & {$\phantom{1}9.03(3)$} & 
{$0.490(2)$} &
{$0.791(9)$} & 
{$1.90(4)$}\\
\noalign{\vskip0.5ex}
{$6.2$} & {$11.63(2)$} &
{$0.500(2)$} &
{$0.807(8)$} & 
{$1.90(4)$} \\
\noalign{\vskip0.5ex}
\hline
\end{tabular}
\vspace{-0.5cm}
\end{table}

We finally mention that the one-loop formula
\begin{equation}
  M/m_{\lat}=(2b_0g_0^2)^{-4/11}
  \left\{1-0.12\times g_0^2+\ldots\right\}
\end{equation}
evaluates to $1.81$ and $1.84$ at $\beta=6.0$ and $\beta=6.2$ respectively.
This compares quite well with the non-perturbative result
quoted in table~1. Note, however, that significantly lower values
would be obtained if the expansion was written in terms of 
Parisi's boosted bare coupling.

\section{CONCLUSIONS}

In this work we have demonstrated that scale-dependent renormalization
factors can be calculated non-perturbatively without compromising
approximations.  The methods that we have employed are completely
general and are hence expected to be useful in other contexts
as well.

Quark mass values are usually quoted in the $\msbar$ scheme 
of dimensional regularization 
at a normalization mass $\mu=1$ GeV or $\mu=2$ GeV.
Once a non-perturbative 
solution of the theory becomes possible, this convention
is not entirely satisfactory, because the $\msbar$ scheme is only
meaningful to any finite order of perturbation theory.
The renormalization group invariant masses, on the other hand,
are non-perturbatively defined and scheme-independent.
These quark masses are hence more quotable than the $\msbar$
masses and we would like to recommend their use in future studies.

Previous computations of the $\Lambda$-parameter in LQCD have shown that
this is a rather elusive quantity. The basic problem is that it refers
to the high-energy limit of the continuum theory which is not
readily accessible on the lattice. 
Applying our recursive method we have now
been able to overcome this difficulty and to obtain the 
$\Lambda$-parameter with completely controlled errors
(apart from quenching which remains to be the major
limitation of present-day LQCD).

\vskip1ex
This work is part of the ALPHA collaboration research programme.
We thank DESY for allocating computer time to this project.


\end{document}